\documentclass[aip,preprint]{revtex4}
\usepackage{amsmath}
\usepackage{amsfonts}
\usepackage{amssymb}
\usepackage{amsthm}
\usepackage{newlfont}
\usepackage{setspace}
\usepackage{graphicx,color}
\usepackage{natbib}
\usepackage{hyperref}
\newcommand{\pt}[1]{\frac{\partial#1}{\partial t}}

\newcommand{\vvec}{\mathbf{v}}
\newcommand{\Bvec}{\mathbf{B}}
\newcommand{\Avec}{\mathbf{A}}

\newcommand{\Jvec}{\mathbf{J}}

\renewcommand*{\v}[1]{\hbox{\bfseries #1}}
\renewcommand*{\t}[1]{\hbox{\sffamily\bfseries #1}}
\begin{document}

\begin{titlepage}
  \title{Overview of HiFi -- implicit spectral element code framework
    for multi-fluid plasma applications.}
  \author{V. S. Lukin\footnote{email: vlukin1@mailaps.org.  Any opinion, findings, and conclusions or recommendations expressed in this material are those of the authors and do not necessarily reflect the views of the National Science Foundation.}}
  \affiliation{National Science Foundation,
    Arlington, VA 22230.}
 \author{A. H. Glasser}
  \affiliation{Fusion Theory \& Computation, Inc.,
    Kingston, WA, 98346.}
 \author{W. Lowrie}
  \affiliation{Applied Research Associated, Inc., USA}
 \author{E. T. Meier}
  \affiliation{College of William and Mary, Williamsburg, VA 23187}

\begin{abstract}
  An overview of the algorithm and a sampling of plasma applications
  of the implicit, adaptive high order finite (spectral) element
  modeling framework, HiFi, is presented.  The distinguishing
  capabilities of the HiFi code include adaptive spectral element
  spatial representation with flexible geometry, highly parallelizable
  implicit time advance, and general flux-source form of the partial
  differential equations and boundary conditions that can be
  implemented in its framework.  Early algorithm development and
  extensive verification studies of the two-dimensional version of the
  code, known as SEL, have been previously described [A.H.~Glasser \&
  X.Z. Tang, Comp. Phys. Comm., 164 (2004); V.S.~Lukin, Ph.D. thesis,
  Princeton University (2008)].  Here, substantial algorithmic
  improvements and extensions are presented together with examples of
  recent two- and three- dimensional applications of the HiFi
  framework.  These include a Cartesian two-dimensional incompressible
  magnetohydrodynamic simulation of low dissipation magnetic
  reconnection in a large system, a two-dimensional axisymmetric
  simulation of self-similar compression of a magnetic plasma
  confinement configuration, and a three-dimensional Hall MHD
  simulation of spheromak tilting and relaxation.  Some
  planned efforts to further improve and expand the capabilities of
  the HiFi modeling framework are discussed.
\end{abstract}
\maketitle

\end{titlepage}
\section{Introduction}
\label{sec_Intro}

In computational physics community, there is a large number of
existing modeling codes and ongoing development efforts aimed at
efficiently and accurately solving some particular set of partial
differential equations (PDEs) on two-dimensional (2D) or
three-dimensional (3D) grids.  Often, such codes are developed with
the goal of solving a particular physical or engineering problem, and
therefore also assume a particular geometric domain shape.  Fully
periodic pseudospectral turbulence codes
(e.g. Refs.~\cite{Muller00,Perez09}) and toroidally periodic tokamak
modeling codes (e.g. Ref.~\cite{Sovinec04}) are prime examples of such
development efforts.  While being generally effective in solving the
problems they were designed for, such codes are difficult or
impossible to adapt to model closely-related but geometrically
different systems.  Similarly, knowledge of the software and
significant additional code development effort is usually necessary to
modify the system of PDEs under investigation.

On the other hand, there are industry-supported user-friendly software
packages for solving general classes of PDE systems on general and
complex geometric domains (e.g. COMSOL Multiphysics\cite{COMSOL}).
However, these packages are either proprietary and unavailable to the
research community, or extremely inefficient in solving large problems
on modern massively parallel computing systems.

As the demand for computational modeling to simulate existing and
planned scientific experiments, and the need to help understand
fundamental physics of complex dynamical systems grows, the void
between the two types of modeling codes described above has become apparent.  This manuscript describes the two-
and three-dimensional open-source modeling framework called HiFi, which attempts to
partially fill this void for a large class of PDEs that can be written
in the so-called flux-source form:
\begin{equation}
\pt{Q} + \nabla\cdot\vec{F} = S,
\label{eq:primFS}
\end{equation}
where $Q$, $\vec{F}$, and $S$ are functions of time, space, and the
primitive dependent variables, as described below.  Most, if not all,
fluid plasma models can be cast in this form.  (We note that early
algorithm development of the 2D version of the HiFi framework, also
known as SEL, has been described previously by Glasser \& Tang
(2004)\cite{Glasser04}.)  The HiFi framework has been in use for several years and a brief description, with references, of recent modeling studies that have utilized HiFi can be found at \href{url}{http://hifi-framework.webnode.com/hifi-framework/}.

In HiFi, spectral element spatial
discretization\cite{Karniadakis99,Deville02} is used and
Eq.~(\ref{eq:primFS}) is solved in the weak Galerkin form.  The HiFi
framework makes use of implicit time-advance, and is therefore most
beneficial for problems where dynamical time-scales of interest are
much longer than the time it would take the fastest wave to cross the
smallest spatial scale being modeled.  We use the publicly available
PETSc library\cite{petsc} to solve the large linear systems that arise
during the implicit time-advance.  This library is continuously
supported and updated, and allows easy access to other externally
developed direct and iterative linear solvers.  All the main features
of the code are available both in 2D and 3D versions.  The description
presented below will assume 3D spatial representation; and, unless
noted otherwise, it is implied that the same feature is available in
the 2D version.  Extensive verification studies of the 2D version of
the code have been conducted by Lukin~\cite{Lukin08} and later
continued by Meier~\cite{Meier11}.  Verification studies of the 3D
version of HiFi have been performed and reported by
Lowrie~\cite{Lowrie11b}.

In Section~\ref{sec_FluxSource}, we describe the flux-source form
given by Eq.~(\ref{eq:primFS}) in its most general formulation allowed
by HiFi.  In Section~\ref{sec_SpaceDiscr}, the spectral element
spatial discretization and the mapping between the logical space,
where the numerical integration is done, and the physical space, in
which the PDEs are expressed, is presented.
Section~\ref{sec_TimeDiscr} describes the temporal advance options
available in HiFi, as well as the techniques we use to accelerate the
parallel solution of large linear systems resulting from the implicit
formulation.  The boundary condition options available in HiFi are
listed in Section~\ref{sec_BC}.  Additional features and the user
interface provided in HiFi are described in
Section~\ref{sec_UserInterface}.  Results of several 2D and
3D applications are presented in Section~\ref{sec_Appl}.  Summary and future
development plans are presented in Section~\ref{sec_Summary}.
\section{General flux-source formulation}
\label{sec_FluxSource}

Any system of coupled PDEs to be evolved in time by HiFi has to be
expressed in the following general flux-source form as some $M$
number of PDEs of $M$ primary dependent variables
$\left\{U^i(\vec{x})\right\}_{i=1,M}$:
\begin{equation}
\left\{\pt{Q^k} + \nabla \cdot \vec{F}^k = S^k\right\}_{k=1,M}
\label{eq:FS}
\end{equation}
\begin{eqnarray*}
Q^k &\equiv& \sum_{i=1,M}\left[ A^{ki}(\vec{x})
+ \vec{B}^{ki}(\vec{x})\cdot\nabla\right]U^i \\
\vec{F}^k &=&
\vec{F}^k(t,\vec{x},\{U^i\}_{i=1,M},\{\nabla_{\vec{x}}U^i\}_{i=1,M}) \\
S^k &=&
S^k(t,\vec{x},\{U^i\}_{i=1,M},\{\nabla_{\vec{x}}U^i\}_{i=1,M}),
\end{eqnarray*}
where $A^{ki}$, $\vec{B}^{ki}$, $\vec{F}^k$, and $S^k$ are arbitrary
differentiable functions of the given variables and $\vec{x} =
(x,y,z)$ denotes a point vector in the physical metric space $\cal{X}$
in which PDEs are expressed (such as Cartesian, cylindrical, or any
other well-defined coordinate system chosen by the user).  In order to
show how this general form is discretized over any logically cubic
domain $\Xi$, we consider a single PDE of the form of
Eq.~(\ref{eq:FS}) and drop the superscript $k$.  The extension to any
$M$ number of PDEs is straightforward.

In any curvilinear metric space $\Xi$, such that $\vec{\xi} =
(\xi,\eta,\phi)$ are the coordinates of $\Xi$ and ${\cal
  J}(\xi,\eta,\phi) \equiv
(\nabla{z}\cdot\nabla{x}\times\nabla{y})(\nabla{\phi}\cdot\nabla{\xi}\times\nabla{\eta})^{-1}$
is the Jacobian of the transformation from $\cal{X}$ to $\Xi$, it
follows from Eq.~(\ref{eq:FS}) that:
\begin{equation}
    {\cal J}\pt{Q} +
    \frac{\partial}{\partial\xi^i}({\cal J}\vec{F}\cdot\nabla\xi^i)
    ={\cal J}S.
    \label{eq:curvFS}
\end{equation}
(Note, in Eq.~(\ref{eq:curvFS}) and everywhere below we assume the
usual Einstein summation convention.)  Assume that
$x^j=x^j(\vec{\xi})$, for $j=1,3$ is known.
In order to be able to evaluate Eq.~(\ref{eq:curvFS}), it is
necessary to know the coordinate transformation $\nabla\xi^i =
({\partial\xi^i}/{\partial x^j})\nabla x^j$, where expressions
$({\partial\xi^i}/{\partial x^j})$ have to be evaluated in $\Xi$.  We
compute the transformation between $({\partial x^j}/{\partial\xi^i})$
and $({\partial\xi^i}/{\partial x^j})$ under the assumption that $\cal
J$ is non-singular at any location in $\Xi$ where
Eq.~(\ref{eq:curvFS}) is to be evaluated.

Having the coordinate transformations at hand, the rest of the
computations are done in the $\Xi$ metric space.  We call $\Xi$ the
logical space, as the computational domain in $\Xi$ is a cube
$\left(\xi,\eta,\phi)\in([0,1]\times[0,1]\times[0,1]\right)$ with grid
distributed uniformly in $\xi$, $\eta$ and $\phi$.  A mappings $({\cal
  M}: \Xi \rightarrow {\cal X})$ then allows the computational domain
in the physical space to have an arbitrary shape and curvature of the
grid, as long as its topology can be reproduced by identifying
corresponding edges of a structured cube grid.

\section{Spatial discretization}
\label{sec_SpaceDiscr}

The computational domain in HiFi is spatially discretized using the
method of spectral/(hp) elements. (For in-depth discussion on
numerical properties of spectral element discretizations see, for
example, Karniadakis \& Sherwin (1999)\cite{Karniadakis99} and
Deville, Fischer, and Mund (2002)\cite{Deville02} and references
therein.)  Spectral element (or similarly high order finite element)
representation combines the flexibility of an adaptable grid that can
be shaped to fit any given physical domain, parallelization by domain
decomposition, and the exponential spatial convergence, low artificial
wave dispersion and dissipation of purely spectral codes.  Its basic
premise is to have a relatively coarse grid of elements with separate
high order polynomial expansions within each element.  Thus, each
basis function of the overall expansion is identically zero in all but
one or at most several neighboring elements.  The exact set of basis
functions and their coupling across the element boundaries can vary.
For example, among the codes presently employed or being developed in
the MagnetoHydroDynamics (MHD) community, M3D-C1 code\cite{Jardin05}
uses a set of $C^1$-continuous finite elements which are constrained
to be differentiable as well continuous across the element boundaries,
while NIMROD code\cite{Sovinec04} uses a set of $C^0$-continuous
finite elements which only guarantee the continuity of the solution,
but not of its gradients across the element boundaries.

\begin{figure}[htp]
  \center{\includegraphics[width=12cm]{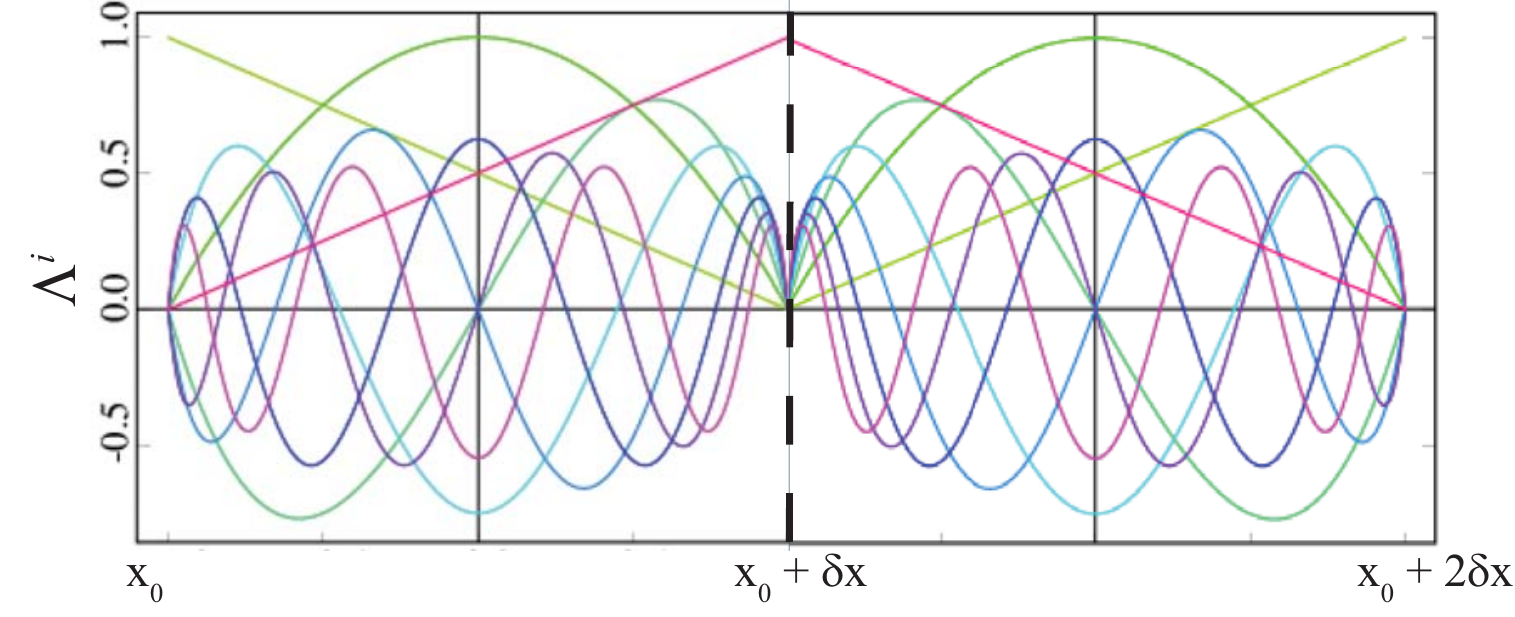}}
  \caption{A one-dimensional illustration of spectral element basis
    functions $\Lambda^i$ used in HiFi.  Shown are two neighboring
    cells with Jacobi polynomial $\{\Lambda^i\}_{i:0,n_p=8}$ basis
    functions in each cell: $\Lambda^0 = (1-\bar{x})/2$,
    $\Lambda^{n_p} = (1+\bar{x})/2$, and $\Lambda^i =
    (1-\bar{x}^2)P_i^{(1,1)}(\bar{x})$, for $i=1,n_p-1$.  In these
    definitions, $\bar{x}\in [-1,1]$ is renormalized from $x\in [x_0 +
    n\delta x, x_0 + (n+1)\delta x]$.  Note that $\Lambda^{n_p}$ from
    a cell on the left is joined with $\Lambda^0$ of the cell on the
    right to form a single basis function to insure continuity, while
    all other basis functions vanish at $x = x_0 + \delta x$.}
  \label{fig:SpectrEl}
\end{figure}

The set of basis functions presently implemented in HiFi is the
$C^0$-continuous set of spectral elements $\{\Lambda^i\}$ given by
Jacobi polynomials. (See Figure~\ref{fig:SpectrEl}), where all but
the linear basis functions identically vanish at the element
boundaries.  The linear basis functions are the only ones that provide
the continuity of the solution and the coupling between the elements
in each direction.  Representation in $\xi$, $\eta$ and $\phi$
directions of the logical grid described above is done separately with
the complete basis of 3D functions formed by the set of non-zero
Cartesian products of three unidirectional basis functions
$\alpha^{n}(\xi,\eta,\phi) =
\Lambda^{i}(\xi)\Lambda^{j}(\eta)\Lambda^{k}(\phi)$.

Any physical dependent variable $U(t,\vec{x}(\vec{\xi}))$ is expanded
in $\alpha^i(\vec{\xi})$ and time-dependent amplitudes $u_i(t)$:
\begin{eqnarray}
  U(t,\vec{x}) &=& u_i(t)\alpha^i(\vec{\xi}) \\
  U_{x^k}(t,\vec{x}) &=& u_i(t)
\frac{\partial\alpha^i}{\partial\xi^l}
\frac{\partial\xi^l}{\partial x^k}.
\end{eqnarray}
We note that $x^k(\vec{\xi})$ can be represented similarly as:
\begin{equation}
  x^k(\vec{\xi}) = x^k_i\alpha^i(\vec{\xi}).
\end{equation}
Thus, if at some time $t_0$ during a simulation it becomes desirable
to move the calculation from a grid in the physical space represented
by a mapping $(\cal M: \Xi \rightarrow \cal X)$ to a new grid
represented by a new mapping $(\cal M'= \cal L \cal M: \Xi \rightarrow
\cal X)$, where $\cal L$ is some mapping $(\cal L:\Xi \rightarrow
\Xi)$; $U(t_0,\vec{x})$ and $\vec{x}(\vec{\xi})$ would all be
remapped in the same manner.

Observe that Eq.~(\ref{eq:curvFS}) can be rewritten as:
\begin{equation}
  {\cal J}\pt{Q} + \frac{\partial}{\partial\xi^j}
  \left[F_{x^i}{\cal J}
    \frac{\partial\xi^j}{\partial x^i}\right]= {\cal J}S,
    \label{eq:curvFS_exp}
\end{equation}
where $\{F_{x^i}\equiv\vec{F}\cdot\nabla x^i\}_{i=1,3}$ are the
components of the flux of $U$ in the physical space $\cal{X}$.
Reformulating Eq.~(\ref{eq:curvFS_exp}) in the weak form, we have:
\begin{eqnarray}
  \left\{
    \mathbb{M}^{ji}\dot{u_i} \right. &\equiv& \int{\cal J}dV
  \alpha^j\left(A\alpha^i +
    \vec{B}\cdot\nabla\alpha^i\right)\dot{u_i}
  \nonumber \\  &=&
  \int {\cal J}dV\left[S\alpha^j + F_{x^i}
    \left(\frac{\partial\xi^k}{\partial x^i}
      \frac{\partial\alpha^j}{\partial\xi^k}\right)\right] + boundary
  \nonumber \\ &\equiv& \left.
    r^j\left(t,\{u_k\}_{k=1,N}\right)\right\}_{j=1,N},
  \label{eq:Discr}
\end{eqnarray}
where $dV\equiv d\xi~d\eta~d\phi$ and $N$ is the size of the spectral
element basis and therefore is the number of degrees of freedom in
this time-dependent vector equation. (For a system of $M$ PDEs
  on a logical grid with $n_x, n_y, n_z$ elements in $x$-, $y$-, and
  $z$-directions, respectively, and polynomial basis expansion up to
  the $n_p$-th order, the total number of degrees of freedom is $N =
  M*n_x*n_y*n_z*n_p^3$.)

With the derivation above, we have shown how the generalized
flux-source formulation allows for advancing spatially discretized set
of PDEs in an arbitrary logically cubic domain, while the physical
equations can be specified in an unrelated coordinate system most
convenient for one's particular application.  We note that fluxes
$F_x$, $F_y$, $F_z$ and source $S$, together with $A(\vec{x})$ and
$\vec{B}(\vec{x})$, completely specify the PDEs for any given problem,
and the coordinate transformation map $\vec{x}(\vec{\xi})$ specifies
its geometry; with these as input, Eq.~(\ref{eq:Discr}) contains all
necessary information about HiFi's spatial discretization to have the
solution advanced in time.  Such separation of physics, geometry and
solution algorithm is the key to the structural organization of the
HiFi framework.

\section{Adaptive temporal advance algorithm}
\label{sec_TimeDiscr}

The implicit temporal advance in HiFi is accomplished by the
Newton-Krylov iterative method\cite{Glasser04,Lukin08}.  However, like
the rest of the framework, the time-advance module of HiFi is designed
to be easily modifiable for any number of particular
time-discretization schemes.  The principle time-dependent equation to
be solved is Eq.~(\ref{eq:Discr}), which can be written as a vector
equation:
\begin{equation}
\mathbb{M}\dot{\mathbf{u}} = \mathbf{r}(t,\mathbf{u}).
\label{eq:Time}
\end{equation}
Presently, two well known algorithms are implemented to solve
Eq.~(\ref{eq:Time}): the $\Theta$-scheme, with an adjustable
time-centering parameter $\theta$; and a $2^{nd}$ order backward
differencing formula (BDF2)\cite{Bank85}.  Below, we briefly outline
each of the time-discretization schemes.  We then describe the
implementation of the Newton-Krylov iterative advance itself and the
adaptive time-stepping algorithm.

\subsection{$\Theta$-scheme:}
\label{ss_theta}

Equation~(\ref{eq:Time}) is discretized as
\begin{equation}
\mathbb{M}\left(\frac{\mathbf{u}^{n+1} - \mathbf{u}^n}{h}\right) =
\theta\mathbf{r}\left(t^{n+1},\mathbf{u}^{n+1}\right) + (1 -
\theta)\mathbf{r}\left(t^n,\mathbf{u}^n\right),
\label{eq:Theta}
\end{equation}
where $h \equiv \delta t^{n+1} = t^{n+1} - t^n$ is the size of the
$(n+1)$-st time-step. With $\theta = .5$, the $\Theta$-scheme is known
as the Crank-Nicholson method and is an implicit second order
non-dissipative time-discretization method.  All of the application
examples presented in Section~\ref{sec_Appl} advanced PDEs describing
appropriate physical systems with the Crank-Nicholson method.
However, with $\theta$ as a run-time input parameter, both $\theta =
0$ explicit and $\theta = 1$ first order dissipative implicit methods
can also be used for purposes of testing novel PDE implementations.

In order to solve Eq.~(\ref{eq:Theta}) for $\mathbf{u}^{n+1}$ by
Newton's iteration, an initial guess is set to
$\mathbf{u}_0^{n+1}\equiv\mathbf{u}^n$, the change in the solution
being sought is denoted by $\delta\mathbf{u}_i\equiv
\mathbf{u}_{i+1}^{n+1} - \mathbf{u}_i^{n+1}$, the residual
$\mathbf{R}$ is defined as
\begin{eqnarray}
  \mathbf{R}\left(\mathbf{u}_i^{n+1}\right)&\equiv&
  \mathbb{M}\delta\mathbf{u}_i \nonumber \\
  &-& h \left[\theta\mathbf{r}\left(t^{n+1},\mathbf{u}_i^{n+1}\right)
    + (1 - \theta)\mathbf{r}\left(t^n,\mathbf{u}^n\right)\right],
  \label{eq:Theta_res}
\end{eqnarray}
and the Jacobian of the iteration is defined as
\begin{equation}
\mathbb{J}^{ij} \equiv \mathbb{M}^{ij} - h\theta\left\{\frac{\partial r^i}
{\partial u_j}\right\}_{t=t^{n+1}, \mathbf{u}=\mathbf{u}^n}.
\label{eq:Theta_jac}
\end{equation}

\subsection{BDF2 scheme:}
\label{ss_bdf}

Equation~(\ref{eq:Time}) is discretized as
\begin{equation}
\mathbb{M}\left(\frac{\mathbf{u}^{n+1} - a\mathbf{u}^n +
b\mathbf{u}^{n-1}}{h}\right) = \mathbf{r}^{n+1},
\label{eq:bdf}
\end{equation}
where
\begin{equation*}
a \equiv \frac{\left(\delta t^n + \delta t^{n+1}\right)^2} {\delta
t^n\left(\delta t^n + 2\delta t^{n+1}\right)},
\end{equation*}
\begin{equation*}
b \equiv \frac{\left(\delta t^{n+1}\right)^2}{\delta t^n\left(\delta t^n +
2\delta t^{n+1}\right)},
\end{equation*}
\begin{equation*}
h \equiv \frac{\delta t^{n+1}\left(\delta t^n + \delta t^{n+1}\right)}
{\left(\delta t^n + 2\delta t^{n+1}\right)},
\end{equation*}
$\delta t^n = t^n - t^{n-1}$, and $\delta t^{n+1} = t^{n+1} - t^n$. Here, an
initial guess is set to $\mathbf{u}_0^{n+1}\equiv a\mathbf{u}^n -
b\mathbf{u}^{n-1}$, change in the solution is again
$\delta\mathbf{u}_i\equiv\mathbf{u}_{i+1}^{n+1} - \mathbf{u}_i^{n+1}$, the
residual is defined as
\begin{equation}
\mathbf{R}\left(\mathbf{u}_i^{n+1}\right)\equiv\mathbb{M}\delta\mathbf{u}_i
    - h\mathbf{r}\left(t^{n+1},\mathbf{u}_i^{n+1}\right),
\label{eq:bdf_res}
\end{equation}
and the Jacobian of the iteration is
\begin{equation}
\mathbb{J}^{ij} \equiv \mathbb{M}^{ij} - h\left\{\frac{\partial r^i} {\partial
u_j}\right\}_{t=t^{n+1}, \mathbf{u}=\mathbf{u}^n}.
\label{eq:bdf_jac}
\end{equation}
Like Crank-Nicholson, BDF2 is also a second order time-discretization
method.  However, straightforward analysis of Eq.~(\ref{eq:bdf})
demonstrates that BDF2 damps high time-frequency modes of the
solution, thus providing numerical dissipation in the algorithm.  When
using the BDF2 scheme, we resolve the issue of the first time-step by
making the first time-step with the $\Theta$-scheme, and then taking
the initial condition and the first time-step as the $(n-1)$-st and
the $n$-th values of $\mathbf{u}$, respectively. We also note that
Eqs.~(\ref{eq:bdf})-(\ref{eq:bdf_jac}) explicitly allow for $\delta
t^{n+1}\neq\delta t^n$, which is necessary to have an adaptive
time-stepping algorithm.

Using either of the time-discretization schemes described above, time
advance is accomplished by iterating on
\begin{eqnarray}
  \mathbf{R}_i + \mathbb{J}\delta\mathbf{u}_i = 0  &\rightarrow&
  \delta\mathbf{u}_i = - \mathbb{J}^{-1}\mathbf{R}_i \nonumber \\
  &\rightarrow&
  \mathbf{u}_{i+1}^{n+1} = \mathbf{u}_i^{n+1} + \delta\mathbf{u}_i
  \nonumber \\
  i &\Rightarrow& i+1
\label{eq:Newton}
\end{eqnarray}
until the condition $\mathfrak{N}(\mathbf{R}_i)\leq n_{tol}$ is
satisfied, where $\mathfrak{N}$ is the $\mathbb{L}^2$ norm of
$\mathbf{R}_i$ normalized to $\mathbf{R}_0$ and $n_{tol}$ is a
run-time input parameter determining the tolerance of the Newton
iteration convergence.  Once the Newton iteration has converged, the
solution vector is advanced by setting $\mathbf{u}^{n+1} =
\mathbf{u}_{i+1}^{n+1}$.  An advanced non-linear Newton solver
available through PETSc, {\it SNESSolve}\cite{petsc}, is used in the
current HiFi implementation to complete the above cycle.

The Newton iteration procedure includes a non-trivial step of solving
the matrix $\mathbb{J}$, which is an $N\times N$ sparse matrix, where
$N$ is the total number of degrees of freedom.  In fact, $\mathbb{J}$
describes the exact coupling between each of the degrees of freedom at
time $t=t^{n}$. However, due to the $C^0$ nature of the basis
functions employed in HiFi, only ``skeletons'' representing the linear
basis functions (linear in at least one direction) within each cell
are coupled to each other across the cell boundaries. The so-called
static condensation procedure\cite{Karniadakis99}
separates the skeletons from the interiors of the cells and uses
separate local solves for each of the cell's
interiors\cite{Glasser04}.  By doing so, static condensation both
reduces the size of the global matrix to be solved by a factor of
$n_p$ and significantly improves the parallel efficiency of the code.
We note that in order to enable the static condensation algorithm, the
matrix $\left\{\partial r^i/\partial u_j\right\}$ involved in
calculating $\mathbb{J}$ in both Eq.~(\ref{eq:Theta_jac}) and
Eq.~(\ref{eq:bdf_jac}) has to be calculated explicitly by taking
derivatives of Eq.~(\ref{eq:Discr}) with respect to all degrees of
freedom in the system.  This is accomplished by specifying the
analytical derivatives of the fluxes $F_{x^i}$ and sources $S$ with
respect to the evolved physical variables $U$ and their gradient
components $U_{x^i}$. Though somewhat labor-intensive in coding, this
method allows for much greater accuracy of the time-advance algorithm.

An additional method of preconditioning the HiFi linear system is
presently under development.  So-called physics-based preconditioning
(PBP), originally developed by Luis Chac\`{o}n in the context of a
finite volume spatial discretization\cite{Chacon08}, is designed to
achieve near-perfect weak scalability in solving linear systems
resulting from implicit advance of discretized MHD systems over tens
of thousands of processors and beyond.

The remaining global matrix is solved in parallel using the PETSc
libraries\cite{petsc} with the linear solvers available and
appropriate for any given problem.  Choice of any particular solver,
such as direct LU factorization or the flexible Generalized Minimal
Residual (fGMRES) method, is made at run-time and requires no
modifications to the code.  Local solves are accomplished with LAPACK
routines.

We now return to Equation~(\ref{eq:Newton}) and consider what happens
if a time step $\delta t^{n+1}$ taken in Eq.~(\ref{eq:Theta}) for
$\Theta$-scheme or in Eq.~(\ref{eq:bdf}) for BDF2 is either
unnecessarily small, so that Newton iterations converge \emph{too
  quickly}, or so large that \emph{too many} iterations are necessary
for convergence.  Run time input parameters $newt_{max}$ and
$newt_{min}$ define those limits for each particular simulation run.
The automatic adaptivity of the time-step is accomplished by
decreasing $\delta t^{n+1}$ by some fraction $f_{decr} < 1$ and
recalculating the Jacobian whenever Eq.~(\ref{eq:Newton}) has not
converged after $newt_{max}$ Newton iterations.  Conversely, $\delta
t^{n+1}$ is set to $\delta t^{n+1} = f_{incr}\delta t^{n}$, $f_{incr}
> 1$, whenever the Newton iterations of the previous time-step
converged in less than $newt_{min}$ number of iterations.  For
iterative linear solvers such as fGMRES, the number of fGMRES
iterations can be an additional factor in determining whether or not
to increase/decrease the time step.  This simple algorithm has proven
to be very robust and useful in modeling systems that have both long
periods of slow and/or linear evolution and bursts of activity with
very short non-linear dynamical time-scales\cite{Lukin08}.

Additional performance gain has been achieved by re-evaluating the
Jacobian $\mathbb{J}$ only during those time-steps when the number of
Newton iterations $it_N$ taken during the previous time-step was equal
or greater than $newt_{max}$.  However, if $newt_{max} > it_N \ge
newt_{min}$, the Jacobian matrix used during the previous time-step is
re-used without being re-evaluated.  While allowing for significant
gain in performance, particularly during quasi-linear periods of
evolution in any number of non-linear simulations, this technique does
not lead to any deterioration in the accuracy of the computation.
\section{Formulation of boundary conditions}
\label{sec_BC}

As indicated in Eq.~(\ref{eq:Discr}), formulation of boundary
conditions in HiFi is integrated into the overall flux-source form.
All quantities are advanced in time on the boundary and in the
interior of the domain in a single time-step by solving the primary
system of PDEs in the interior together with a separate system of PDEs
describing the boundary conditions on the boundary nodes.  Two classes
of general boundary condition (BC) forms, as well some special cases,
are available in HiFi.

We call one of the BC classes -- the explicit local BC form, where the
solution on the boundary must satisfy a general non-linear
time-dependent equation of the form
\begin{equation}
  \left\{\left[A^{ki}\pt{U^i}
      + \vec{B}^{ki}\cdot\nabla\left(\pt{U^i}\right)\right] = S^k\right\}_{k=1,M}
    \label{eq:BC_local}
\end{equation}
\begin{equation*}
  S^k = S^k\left(t,\hat{n},\vec{x},
    \{U^i,\nabla_{\vec{x}}U^i,\nabla_{\vec{x}\vec{x}}U^i\}_{i=1,M}\right)
\end{equation*}
where $A^{ki} = A^{ki}(\hat{n},\vec{x})$, $\vec{B}^{ki} =
\vec{B}^{ki}(\hat{n},\vec{x})$, and $S^k$ are arbitrary differentiable
functions of the given variables and $\hat{n}$ denotes an outward unit
vector normal to the boundary of the domain.

The other BC class -- the flux BC form, allows users to specify the
desired normal flux $F_n\equiv\vec{F}\cdot\hat{n}$ of a particular
primary dependent variable through the boundary of the domain.  Once
again,
\begin{equation}
  F_n = F_n\left(t,\hat{n},\vec{x},
    \{U^i,\nabla_{\vec{x}}U^i,\nabla_{\vec{x}\vec{x}}U^i\}_{i=1,M}\right)
    \label{eq:BC_flux}
\end{equation}
can be an arbitrary differentiable function of the given variables.

Two special boundary condition options are also available: (1)
periodic BC's in any or all directions can be imposed on the full
system, or on specific dependent variables; (2) cylindrical BC can be
imposed on the system, such that for any $\eta_0\in[0,1]$, all points
$(\xi,\eta,\phi)\in(\xi,\eta_0,0)$ in the 3D logical space are
identified together. (In the 2D implementation, there is an equivalent
polar BC option, where all points $(\xi,\eta)\in(0,\eta)$ in the 2D
logical space are identified together.)
\section{User interface and additional features}
\label{sec_UserInterface}

Making use of the generic implementation of the primary PDE system,
boundary conditions, and the physical domain shape -- the HiFi
user interface consists of a standardized set of subroutines collected
into a {\it physics} template file.  Within the template file, the
user has the freedom
\begin{enumerate}
\item to specify the functional forms that would uniquely determine
  Eq.~(\ref{eq:FS});
\item to choose the class of boundary conditions separately on each
  face for each dependent variable and subsequently specify the
  necessary functional forms to uniquely determine either
  Eq.~(\ref{eq:BC_local}) or Eq.~(\ref{eq:BC_flux});
\item to specify the initial map between the logical and physical
  spaces;
\item to specify the initial conditions, as well as the set of
  user-desired input variables for the problem at hand.
\end{enumerate}
The rest of the HiFi algorithm is separated and compiled into a
library, that can be used with any {\it physics} application file constructed
according to the template.  We note that as long as the set of
specified primary PDEs and boundary conditions has a unique solution,
any of the free functions provided in
Eqs.~(\ref{eq:FS},\ref{eq:BC_local},\ref{eq:BC_flux}) can also be set
to zero.

One of the most attractive additional features of the HiFi framework
is grid adaptation.  There are a number of strategies and approaches
that have been attempted in the computational physics community to
enable accurate and efficient grid adaptation for solving
initial-value problems with multi-scale spatial behavior.  They can be
generally divided into two groups: adaptive mesh refinement (AMR),
where parts of the grid with insufficient resolution are refined by
effectively subdividing the existing grid
cells\cite{Berger84,Berger89}; and dynamic
Arbitrary-Lagrangian-Eulerian (ALE)
techniques\cite{Hirt74,Jardin78,Kershaw98} and/or variational
principle based harmonic grid
generation\cite{Brackbill86,Brackbill93,Knupp95}, where an evolving
mapping between some logical grid of a fixed size and the physical
domain provides the necessary adaptation.  Algorithms that combine the
two approaches are also being developed\cite{Anderson04}.  While each
of the methods has its advantages and drawbacks in flexibility,
accuracy and parallel efficiency, we have chosen to pursue a harmonic
grid generation method which appears to be highly accurate, relatively
flexible and does not in any way inhibit the parallel efficiency of
the HiFi framework.  We have collaborated with Liseikin\cite{Liseikin03} in
the development of such grid generation algorithm capable of finding
an optimal mapping $\mathcal{M}$ between the logical domain $\Xi$ and
given physical domain $\cal X$.  The details of the HiFi adaptive grid
implementation and verification studies have been reported by
Lukin\cite{Lukin08} and will be further described in a follow-up
manuscript.

Another useful feature of HiFi is the ability to restart a simulation
from a previously generated check-point data file, while either
increasing or decreasing the overall resolution of the restarted
simulation.  Furthermore, such previously generated data may come from
a solution of an entirely different set of PDEs with different
dependent variables: for example, the user can read in the solution of
some anisotropic heat conduction equation to initialize the
temperature in a compressible MHD simulation.

We take advantage of the parallel HDF5 libraries\cite{hdf5} for the
check-point data input and output (IO).  In order to visualize or
extract quantitative physically meaningful results from the computed
data, the check-point files are additionally post-processed.  Parallel
post-processing is presently available for the 3D data.  HiFi's
primary visualization tool, particularly in 3D, is the publicly
available VisIt Visualization Tool\cite{visit}.
\section{Sample Plasma Applications}
\label{sec_Appl}

A number of publications reporting results obtained with various
applications of the HiFi framework are already available.  HiFi
has been used to study idealized physical
systems\cite{Lukin09,Lukin11}, to conduct realistic simulations with
validation against experimental data\cite{Gray10,Cothran10}, to study
numerical properties of the $C^0$ spectral element spatial
discretization\cite{Meier10,Lowrie11a}, and to develop and test new
numerical methods, in particular, for accurate formulation of ``open''
boundary conditions in mixed hyperbolic-parabolic systems of
PDEs\cite{Meier12}.  Here, we briefly describe several ongoing
applications and test verification problems solving different sets of
PDEs with the 2D and 3D HiFi versions in order to demonstrate the
accuracy and flexibility of the framework.
\subsection{Reduced MHD plasmoid-facilitated magnetic reconnection}

One of the simplest 2D systems of PDEs that describe behavior of a
magnetized plasma is the visco-resistive reduced (incompressible) MHD
system of equations, which is valid in the limit of strongly
magnetized collisional plasma.  Assuming no initial variation in the
out-of-plane $\hat{z}$-component of magnetic field $\Bvec$ and no initial
out-of-plane plasma flow $\vvec$, this system of PDEs can be written
in the flux-source form as follows:
\begin{eqnarray}
\pt{\psi} &+& \nabla\cdot\left(\psi\hat{z}\times\nabla\phi\right) =
\nu j
\label{eq:RMHD_psi} \\
\pt{\omega} &+&
\nabla\cdot\left[\omega\hat{z}\times\nabla\phi -
j\hat{z}\times\nabla\psi - \mu\nabla\omega)\right] = 0
\label{eq:RMHD_omega} \\
\nabla&\cdot&\left[\nabla\psi\right] = j
\label{eq:RMHD_j} \\
\nabla&\cdot&\left[\nabla\phi\right] = \omega,
\label{eq:RMHD_phi}
\end{eqnarray}
where $\psi$ is the magnetic flux function with
$\Bvec=\hat{z}\times\nabla\psi$, $\phi$ is the plasma flow stream
function with $\vvec=\hat{z}\times\nabla\phi$, $\nu$ is isotropic
plasma resistivity and $\mu$ is isotropic kinematic viscosity.  HiFi
implementation of and simulations using
Eqs.~(\ref{eq:RMHD_psi}-\ref{eq:RMHD_phi}) have been reported
previously\cite{Lukin08,Glasser04}.  Here, we present results of a
magnetic reconnection simulation similar to those described by
Lukin\cite{Lukin08}, but with lower dissipation parameters $\nu$ and
$\mu$.

The reduced visco-resistive MHD equations,
Eqs.~(\ref{eq:RMHD_psi}-\ref{eq:RMHD_phi}), are solved in a
rectangular box $(x,y)\in[-L_x,L_x]\times[-L_y,L_y]$.  Periodic
boundary conditions are used in the reconnection outflow
$\hat{x}$-direction, while an ``open'' boundary is assumed in the inflow
$\hat{y}$-direction in order to reduce the effects of the domain
boundary on the reconnection layer.  Here, we define ``open'' boundary
to have zero tangential flow, zero vorticity and constant and uniform
tangential component of magnetic field.  Thus, on the y-boundary,
$\hat{y}\cdot\nabla\phi=0$, $\nabla^2\phi=0$, and
$\hat{y}\cdot\nabla\psi=const$ are the enforced BC. Simulations are
initialized with a Harris equilibrium\cite{Harris62} with an
additional small and localized perturbation: $\psi_0 =
\lambda\ln\left[\cosh(y/\lambda)\right] + \delta\psi$,
$\delta\psi=\epsilon\exp\left[-x^2/(2\lambda)^2\right]
\exp\left[-y^2/(\lambda/2)^2\right]$, where $\lambda$ is the
half-width of the Harris equilibrium and $\epsilon$ is the magnitude
of the perturbation.  Note that the perturbation is localized within
the equilibrium current sheet.

\begin{figure}[htp]
  \center{\includegraphics[width=9cm]{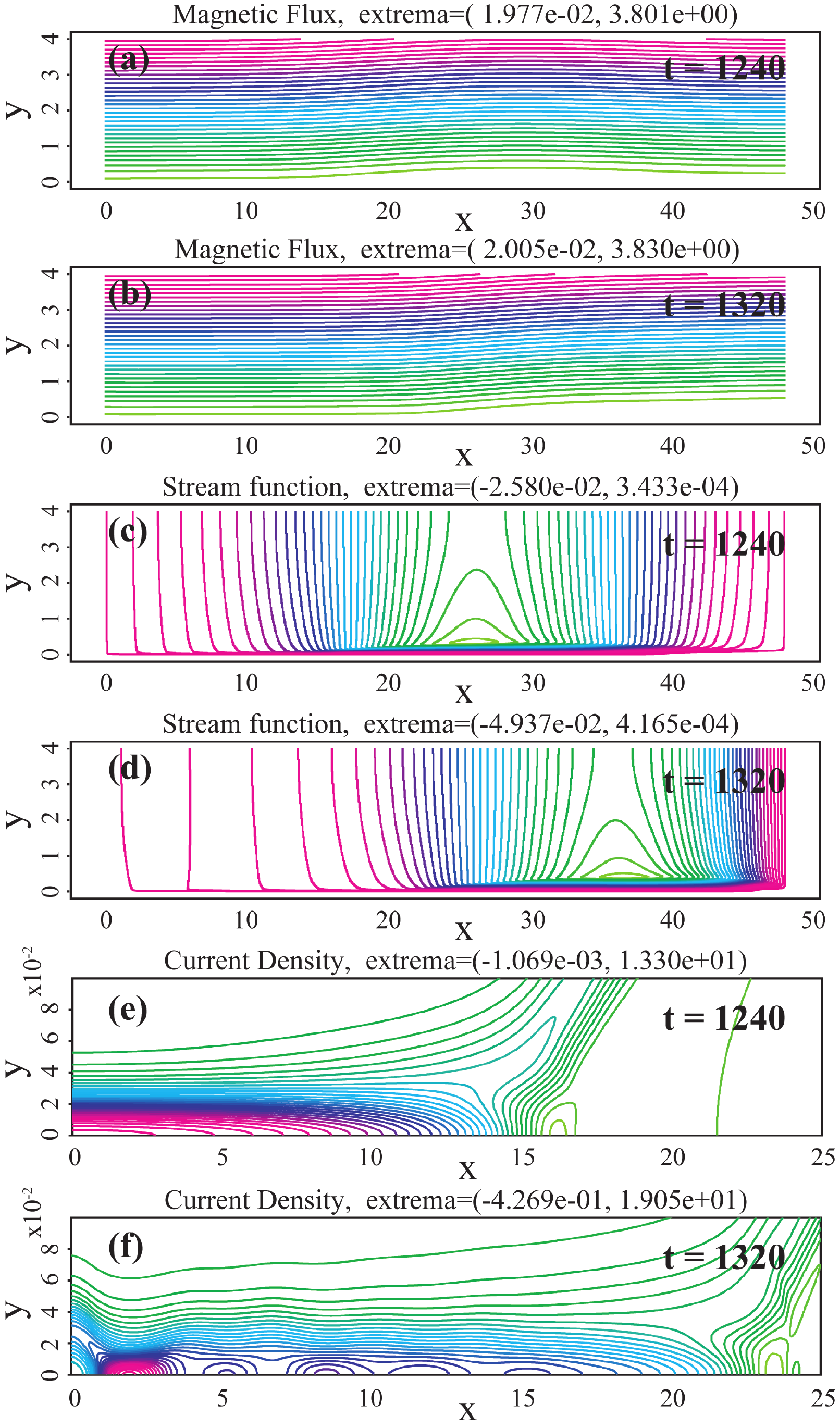}}
  \caption{Contour plots of (a,b) magnetic flux $\psi$, (c,d) stream
    function $\phi$ and (e,f) current density $j$ from a 2D reduced
    MHD magnetic reconnection simulation.  Panels (a-d) show the full
    computational domain, while panels (e,f) show a zoom-in into the
    reconnection region.  Panels (a,c,e) show a single highly
    elongated reconnection region at $t=1240$; while a short time
    later at $t=1320$, panels (b,d,f) show the reconnection region that
    continues to elongate and simultaneously splinters into multiple
    shorter current sheets.  The simulation is conducted with
    resistivity $\nu=10^{-5}$ and viscosity $\mu=10^{-5}$.}
  \label{fig:2DMagRec}
\end{figure}

In order to model the development of a macroscopic resistive current
layer from a local perturbation in a large system, the following
simulation parameters are chosen: $\lambda=.5$, $L_x=48$, $L_y=4$,
$\epsilon=10^{-4}$ and $\nu=\mu=10^{-5}$, where the width of the
initial Harris equilibrium is taken as the effective unit length.
Making use of the symmetries of the initial conditions and those
inherent in Eqs.~(\ref{eq:RMHD_psi}-\ref{eq:RMHD_phi}), simulations
are conducted only in the top-right quarter domain and appropriate
symmetry BC are applied.  No grid adaptation is used in the
simulation.  However, a smooth mapping
$\{\mathcal{M}:(\xi,\eta)\rightarrow
(x,y)=(L_x\xi,L_y[\tanh(\alpha\eta-\alpha)/\tanh(\alpha)+1])\}$
between the logical and physical spaces with $\alpha=2$ provides
computational grid that is highly concentrated near $y=0$, where the
thin resistive reconnection layer shown in Figure~\ref{fig:2DMagRec} forms
during the simulation.

Figure~\ref{fig:2DMagRec} shows contour plots of (a,b) magnetic flux
$\psi$, (c,d) stream function $\phi$ and (e,f) current density $j$
from the simulation on the logical grid of size
$(n_x,n_y,n_p)=(108,48,8)$.  Note that panels (a-d) show the full
computational domain, while panels (e,f) show a zoom-in into the
reconnection region.  It is apparent that results both in panels (a,c,e),
showing a single highly elongated reconnection region at $t=1240$, and
in panels (b,d,f), showing the reconnection region that continues to
elongate and simultaneously splinters into multiple shorter current
sheets at $t=1320$, are very well resolved.

Yet, we have not been able to converge the simulation setup presented
here in spatial resolution.  Decreasing the resolution causes the
reconnection current sheet to splinter earlier, generating multiple
magnetic islands and current sheets.  On the other hand, increasing
the resolution prolongs the single highly elongated current layer
reconnection and expansion until some later time, when it eventually
succumbs to what appears to be the multiple plasmoid instability
described by Loureiro, {\it et al.}\cite{Loureiro07}.  Thus, the
macroscopic behavior of this system is critically influenced by
the level of background noise, determined here by the spatial
resolution.  Similar behavior in semi-collisional Hall MHD magnetic
reconnection simulations has also been previously
observed\cite{Lukin08}.

We note that in a real physical system some level of
background noise is always present and the system size is limited by
the curvature of the global magnetic fields.  Therefore, we expect
that in strongly magnetized collisional plasmas, for any given degree
of collisionality, the length of a macroscopic reconnection region and
the characteristic number of plasmoids, if any, on average contained
within the reconnection region, are determined by the magnitude of the
background noise level relative to the rate of the reconnection region
expansion.
\subsection{FRC compression in visco-resistive MHD}
Another recent application of the HiFi framework is a 2D model of
Magnetized Target Fusion (MTF)\cite{Wurden96,Taccetti03}.  The usual
compressible MHD system of PDEs is solved with 6 dependent variables,
$ (\rho, -A_\phi, p, \rho v_z, \rho v_r, J_\phi)$.  The equations in
cylindrical $r,z$ coordinates are
\begin{eqnarray}
\frac{\partial\rho }{\partial t}
	&+& \nabla\cdot\left(\rho\vvec - D\nabla\rho\right)
        = 0 \\
-\frac{\partial A_{\phi}}{\partial t} &=& v_r B_z - v_z B_r + \eta J_\phi \\
\frac{3}{2}\frac{\partial p}{\partial t}
	&+& \nabla\cdot\left(\frac{5}{2} p\vvec
	- \kappa\cdot\nabla T\right)
        = \eta J_\phi^2 + \pi:\nabla\vvec \\
\frac{\partial(\rho\vvec)}{\partial t}
	&+& \nabla\cdot\left(\rho\vvec\vvec + p\mathbb{I} + \pi\right)
        = (J_{\phi}\hat{\phi})\times\Bvec \\
 J_\phi &=& \frac{A_\phi}{r^2} - \nabla^2 A_\phi,
\end{eqnarray}
where $\Bvec=B_r\hat{r} + B_z\hat{z}=\nabla A_{\phi}\times\nabla\phi$,
$\rho\vvec=\rho v_r\hat{r} + \rho v_z\hat{z}$, $D$ represents
kinematic density diffusion, $\eta$ is resistivity, $\kappa$ is the
anisotropic heat conduction tensor, and $\pi$ is the viscous tensor.
We note that in the absence of $\hat{\phi}$-components of $\Bvec$ and
$\vvec$ in the initial condition, such as in the problem described
below, the symmetries of compressible MHD preserve that property
throughout a simulation.  Thus, we are justified in omitting
$\hat{\phi}$-components $\Bvec$ and $\vvec$ from the above system of
PDEs.

A unique feature of this simulation is the use of a scaled coordinate
system.  The MTF concept involves forming a Field Reversed
Configuration (FRC) in a cylindrical flux conserver and then
compressing it radially by a factor $\sim 10$.  The most efficient way
to model this is to use a grid whose dimensions scale with the motion
of the wall.  We derive equations that allow us to specify this
scaling transformation in the application portion of the code, without
requiring any modification of the larger solver portion of the code.

\begin{figure}[htp]
  \center{\includegraphics[width=12cm]{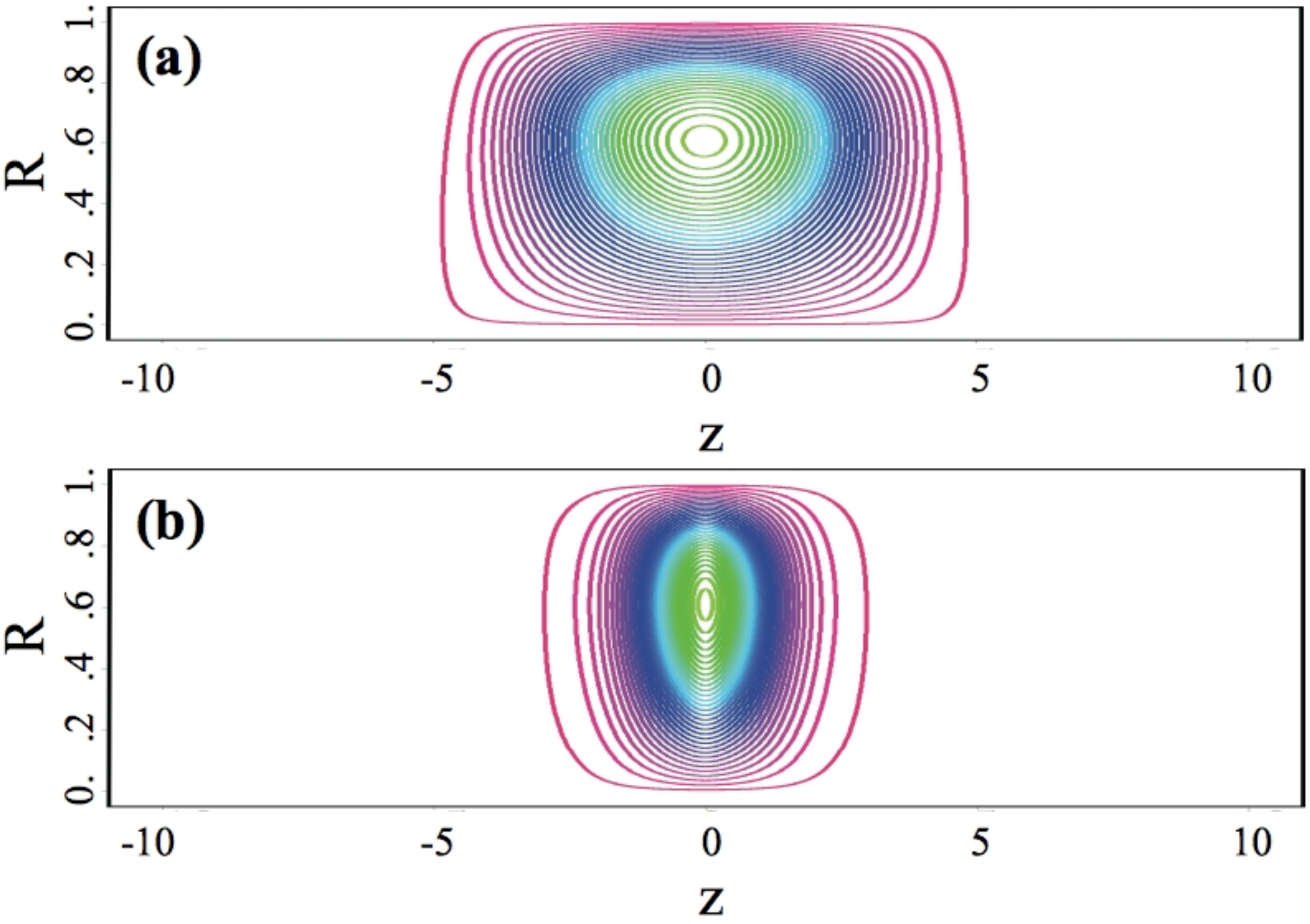}}
  \caption{Contour plot of magnetic vector potential $A_\phi$ {\it
      vs.}  scaled radial variable $R$ and physical axial variable $z$
    before [panel (a)] and after [panel (b)] radial compression by a
    factor of 10.  Note that the FRC experiences axial as well as
    radial compression due to magnetic tension.}
\label{fig:vector_potential}
\end{figure}

\begin{figure}[htp]
  \center{\includegraphics[width=8.00cm]{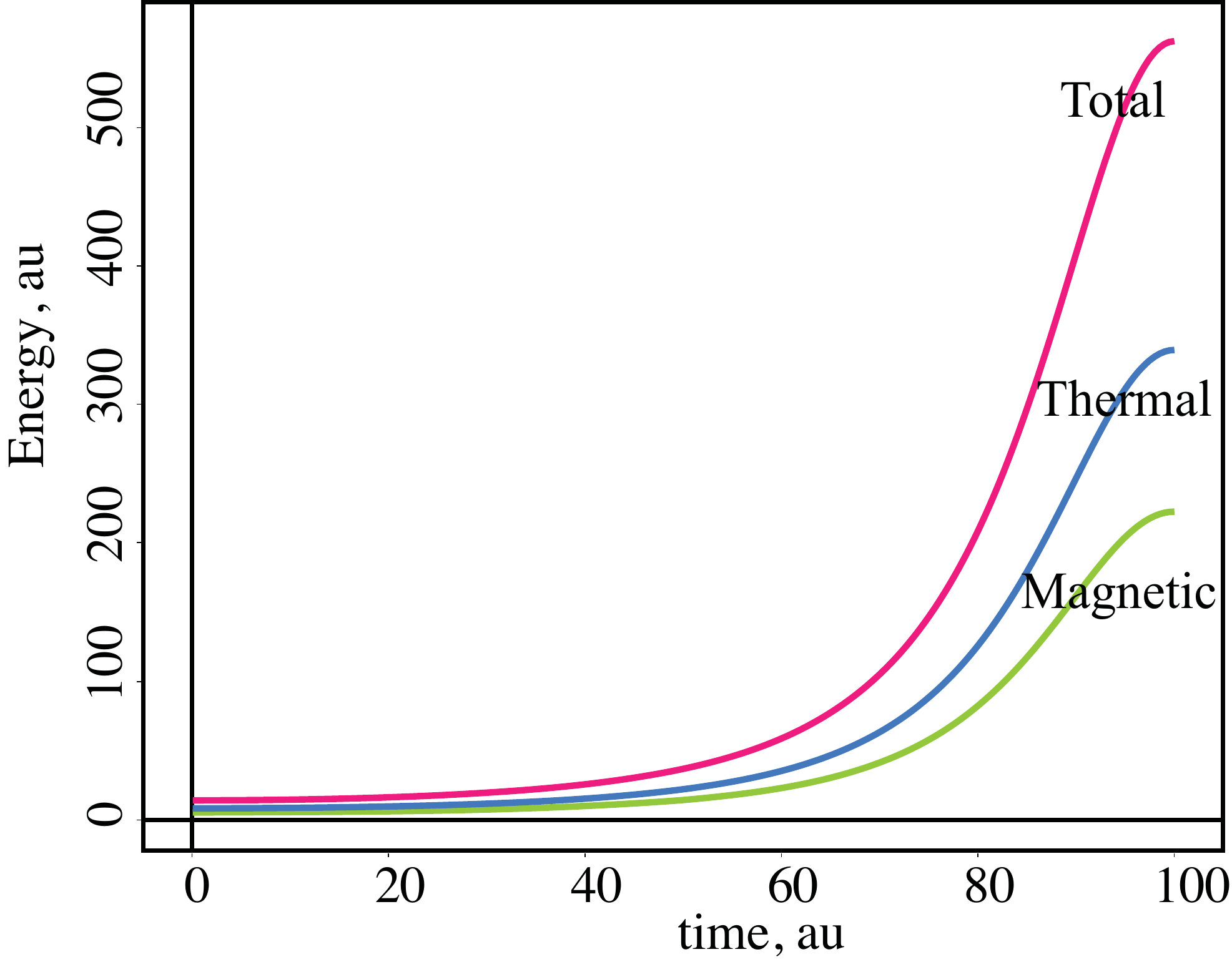}}
  \caption{Volume integrals of total, thermal, and magnetic energy {\it
      vs.} time t during the radial compression.}
\label{fig:energy}
\end{figure}

Let $\v{x}$ and $\v{X}$ denote Cartesian representations of the
physical and scaled position vectors, and let $\t{T}(t)$ represent a
time-dependent scaling transformation, such that
\begin{equation}
\v{x}(\v{X},t) \equiv \t{T}(t) \cdot \v{X}, \quad
\v{X}(\v{x},t) \equiv \t{T}^{-1}(t) \cdot \v{x}
\end{equation}
To compute the function $u(\v{x}(\v{X},t),t) = u(\t{T}(t) \cdot
\v{X},t)$, we use the coordinate transformations
\begin{eqnarray}
\frac{\partial u }{ \partial \v{x}} \Big|_t
   = \frac{\partial }{ \partial \v{x}} \v{X} \cdot
   \frac{\partial u }{ \partial \v{X}} \Big|_t
   = \t{T}^{-1} \cdot \frac{\partial u }{ \partial \v{X}} \Big|_t \\
\frac{\partial u }{ \partial t} \Big|_{\v{x}}
	= \frac{\partial u }{ \partial t} \Big|_{\v{X}}
	- \v{V} \cdot \frac{\partial u }{ \partial \v{x}}, \quad
\v{V} \equiv \frac{\partial \v{x} }{ \partial t}
        \Big|_{\v{X}}
	= \dot{\t{T}} \cdot \v{X}
\end{eqnarray}
with $\dot{\t{T}} \equiv {d\t{T} / dt}$.  A general system of
flux-source equations in physical coordinates of the form
\begin{equation}
A \frac{\partial u }{ \partial t}\Big|_{\v{x}}
	+ \frac{\partial }{ \partial \v{x}} \cdot \v{F}\Big|_t = S
\end{equation}
is then equivalent to the equation in scaled coordinates of the form
\begin{equation}
A \frac{\partial u }{ \partial t}\Big|_{\v{X}}
	+ \frac{\partial }{ \partial \v{X}} \cdot \v{F}'\Big|_t = S'
\end{equation}
with
\begin{equation}
\v{F}' \equiv \v{F} \cdot \t{T}^{-1}, \quad
S' = S + A \left( \dot{\t{T}} \cdot \v{X} \right)
	\cdot \left( \t{T}^{-1} \cdot
	\frac{\partial u }{ \partial \v{X}} \right)
\end{equation}
In the MTF radial compression problem, we define $\t{T}(t)$ to
represent the moving radial wall $r = T(t) R$, with $T(t) =
a\cos(\omega t) + b$, $a = (T_{init} - T_{final})/2$, $b = (T_{init} +
T_{final})/2$, $\omega = \pi/t_{stag}$, with the scaled coordinate
$R\in[0,1]$.  Figs.~\ref{fig:vector_potential}-\ref{fig:energy} show
results of a simulation with $T_{init}=1$, $T_{final}=0.1$, and
$t_{stag}=100$.

The initial conditions in the simulation use a numerical solution of
the Grad-Shafranov equation with the vector potential
$A_{\phi}$ shown in panel (a) of Figure~\ref{fig:vector_potential},
plasma density $\rho\propto p^{1/2}$, and no plasma flow.  The initial
plasma pressure $p$ outside of the FRC magnetic separatrix is set to
be uniform at $0.3\%$ of the peak initial pressure at the center of
the FRC.  Perfectly-conducting, impenetrable, non-slip, thermally
insulating boundary conditions have been imposed at the moving radial
wall, and the system is assumed to be periodic in the axial
$\hat{z}$-direction.

The resulting final magnetic configuration at $t=t_{stag}$ after
radial compression by $T_{init}/T_{final}=10$ is shown in panel (b) of
Figure~\ref{fig:vector_potential}.  Note that in addition to radial
compression, the FRC also experiences axial compression due to magnetic
tension.  Figure~\ref{fig:energy} shows time-traces of total, thermal
and magnetic energy in the system throughout the simulation.  The
force acting to compress the flux-conserver against the thermal and
magnetic back-pressure of the FRC provides the energy source in the
system.  It is clear that most of the energy input goes into the
thermal energy, demonstrating the promise of the MTF method for fusion
applications.
\subsection{3D compressible Hall MHD spheromak tilt study}

An example of 3D HiFi application is a compressible Hall MHD study of
the non-linear dynamics of a tilting spheromak, conducted on a
cylindrical grid solving the following set of normalized PDEs
expressed in the Cartesian coordinate system:
\begin{eqnarray}
\pt{\rho} &+& \nabla\cdot(\rho\vvec_i) = 0 \\
\pt{(\rho\vvec_i)} &+&
\nabla\cdot\left[\rho\vvec_i\vvec_i + p\mathbb{I}
- \bar{\mu}\nabla\vvec_i - \bar{\nu}\nabla\vvec_e\right] \nonumber \\
&=& \Jvec\times\Bvec \\
\pt{\Avec} &=& \vvec_e\times\Bvec + \frac{d_i}{\rho}\nabla p_e
- \bar{\eta}\Jvec - \frac{d_i}{\rho}\bar{\nu}\nabla^2\vvec_e
\label{eq:3DHMHD_OhmsLaw} \\
\nabla&\cdot&\left[(\nabla\cdot\Avec)\mathbb{I}-\nabla\Avec\right] =
\Jvec \\
\frac{3}{2}\pt{p} &+&
\nabla\cdot\left[\frac{5}{2}\left(p_i\vvec_i + p_e\vvec_e\right)
- \bar{\kappa}\nabla T\right] \nonumber \\
&=& \vvec_i\cdot\nabla p_i + \vvec_e\cdot\nabla p_e + \bar{\eta}
|\Jvec|^2
\nonumber \\
&+& \bar{\mu}\nabla\vvec_i:\nabla\vvec_i
+ \bar{\nu}\nabla\vvec_e:\nabla\vvec_e
\end{eqnarray}
where $\Bvec = \nabla\times\Avec$, $\vvec_e = (\rho\vvec_i -
d_i\Jvec)/\rho$, $p = \rho T = p_i + p_e$, $p_e/p_i=\alpha=const$,
$d_i = (c/\omega_{pi})/L_0 = (c/L_0 e) \sqrt{m_i / 4\pi n_0}$,
$\bar{\eta} = (\eta c^2/L_0 B_0)\sqrt{m_i n_0 / 4\pi}$, $\bar{\mu} =
(\mu_i/L_0 B_0)\sqrt{4\pi / m_i n_0}$, $\bar{\nu} = (\mu_e/L_0
B_0)\sqrt{4\pi / m_i n_0}$, $\bar{\kappa} = (\kappa/L_0 B_0)\sqrt{4\pi
  m_i / n_0}$, and \{$\eta,\mu_i,\mu_e,\kappa$\} are some physical
values for resistivity, ion viscosity, electron viscosity and heat
conduction (assumed to be isotropic with $\kappa=\kappa_e=\kappa_i$),
respectively.  Note that all normalizations are determined by the
choices for $L_0$, $B_0$, and $n_0$.

The computational domain is a cylinder of radius $R=L_0$ and length
$L=2L_0$, with the cylindrical BC applied at the cylindrical axis. The
simulation is initialized with a stationary axisymmetric Solov'ev
spheromak equilibrium with uniform normalized pressure and density of
$p=\rho=1$, and $O(10^{-2})$ tilting perturbation in axial ion
velocity $v_{iz}$.  The following perfect conductor
($\hat{n}\times(\partial\Avec/\partial t)=\mathbf{0}$) non-penetrable
($\hat{n}\cdot\vvec_i=0$) energy-conserving BC are imposed: heat
insulator $\hat{n}\cdot\nabla T=0$, perfect slip ion flow
$\hat{n}\cdot\nabla(\hat{n}\times\vvec)=\mathbf{0}$, perfect slip
electron flow $\hat{n}\cdot\nabla\vvec_e=\mathbf{0}$.  Additionally,
$\nabla\cdot\Avec=0$ is imposed to specify the
electro-magnetic gauge BC.

\begin{figure}[htp]
  \center{\includegraphics[width=12cm]{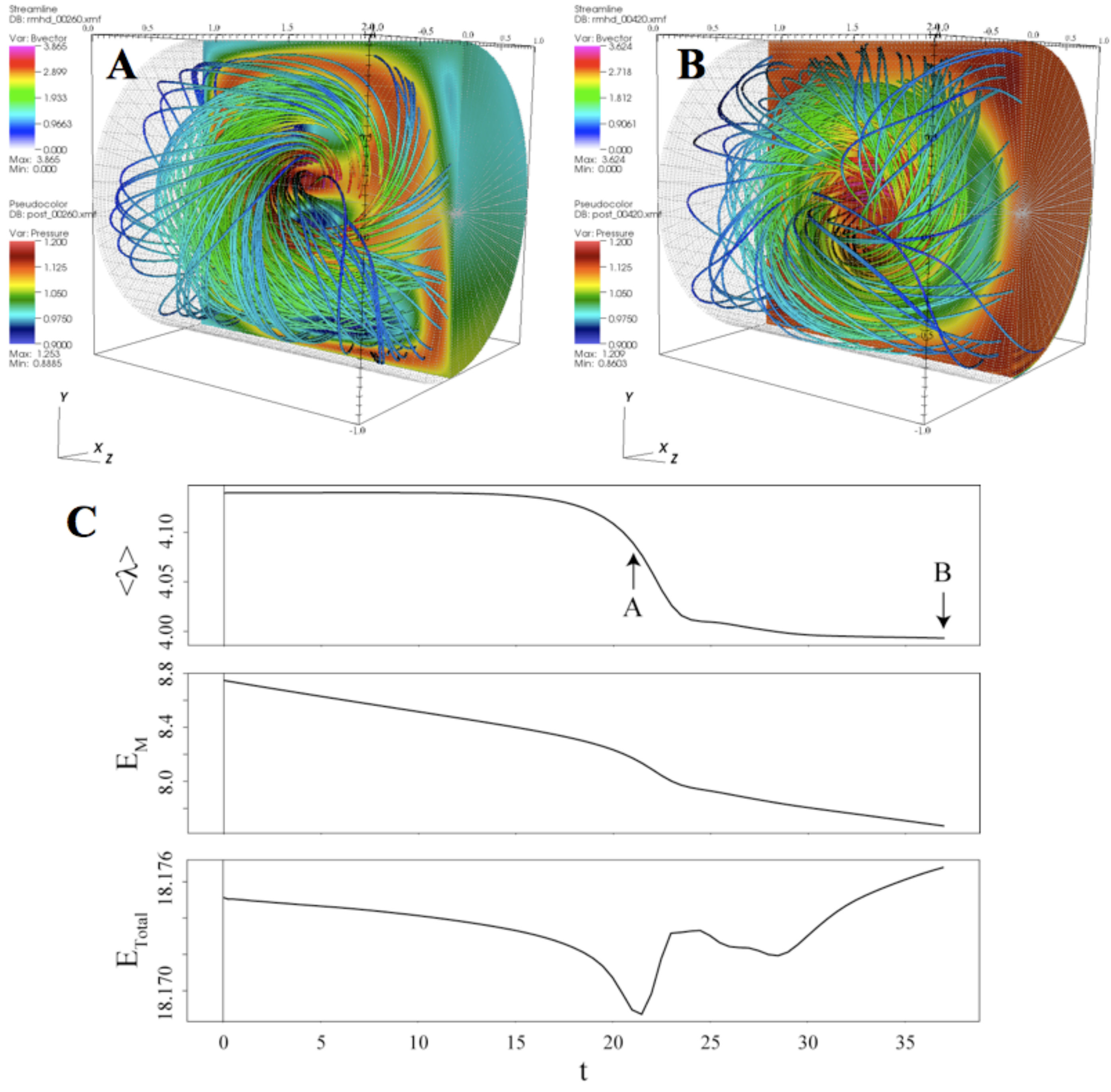}}
  \caption{Frames A and B show magnetic field lines (streamlines
    color-coded by $|\Bvec|$) and plasma pressure (pseudocolor
    cutaway) at two different times during a Hall MHD spheromak tilt
    simulation.  The time-stamps of frames A and B are shown in frame
    C, where the evolution of $<\lambda>\equiv 2E_M/K$ (ratio of
    magnetic energy to magnetic helicity), magnetic energy $E_M$, and
    total energy $E_{Total}$ are shown throughout the simulation.}
  \label{fig:3DHMHD}
\end{figure}
The simulation data shown in Figure~\ref{fig:3DHMHD} was obtained with
the following values for the dimensionless parameters in the PDE
system specified above: $\alpha=0$, $d_i=10^{-1}$, $\bar{\eta}=0$,
$\bar{\mu}=10^{-2}$, $\bar{\nu}=5\times10^{-4}$ and
$\bar{\kappa}=10^{-1}$.  A computational grid of
$(n_r,n_{\phi},n_z,n_p)=(6,6,6,5)$ was used with the grid distributed
uniformly in the radial, angular, and axial directions.

Frames A and B of Figure~\ref{fig:3DHMHD} show streamlines of magnetic
field color-coded by $|\Bvec|$ and pseudocolor cutaway of plasma
pressure $p$ in the midst of the tilting (frame A) and fully relaxed
(frame B).  The time-stamps of frames A and B are shown in frame C,
where the the top panel shows the evolution of $<\lambda>\equiv
2E_M/K$, where $E_M\equiv 1/2\int\Bvec\cdot\Bvec~dV$ is the
magnetic energy and $K\equiv \int\Avec\cdot\Bvec~dV$ is the
magnetic helicity in the system.

We note that in closed systems with low magnetic dissipation, such as
the one considered here, magnetic field is thought to relax to the
lowest available energy state, while its helicity remains
approximately constant\cite{Woltjer58,Taylor74}.  Such relaxed
Woltjer-Taylor states must satisfy
\begin{equation}
\nabla\times\Bvec=\lambda\Bvec,
\label{eq:Taylor}
\end{equation}
where $\lambda$ is a constant.  It is easy to show that in a closed
system where magnetic field satisfies Eq.~(\ref{eq:Taylor}),
$<\lambda>=\lambda$.  In fact, the initial axisymmetric spheromak
state of the simulation can be described by $\lambda=4.138/L_0$, while
the lowest energy Woltjer-Taylor state in the perfectly-conducting
$L:R=2:1$ cylinder has $\lambda=3.978/L_0$ -- and top
panel of frame C in Fig.~\ref{fig:3DHMHD} shows normalized $<\lambda>$
dropping from $4.138$ to $3.978$ as the magnetic fields relax.

The bottom two panels of Fig.~\ref{fig:3DHMHD} show the magnetic
energy $E_M$ and the total energy $E_{Total}$ versus time throughout
the simulation.  We observe that, as expected, the steady loss of
magnetic energy due to the electron viscous term in
Eq.~(\ref{eq:3DHMHD_OhmsLaw}) is accelerated when the spheromak begins
to tilt, and then settles down to a slower rate of decay as the system
approaches the relaxed state.  We measure the total energy of the
system to be conserved to about 5 parts in $10^4$.  It should be
emphasized that we do not evolve total energy as one of the dependent
variables nor use any special techniques designed to conserve energy
in the simulation; the energy conservation is due purely to solving
PDEs with boundary conditions that analytically conserve energy in the
continuous limit.
\section{Summary and Future Development Plans}
\label{sec_Summary}

In this manuscript, we have described the HiFi implicit high order
finite (spectral) element modeling framework for multi-fluid plasma
applications.  The general flux-source form of the PDEs required by
HiFi, the details of the spatial and temporal discretization, the
boundary condition options, as well as the user interface and several
additional features of the framework have been presented.  Several
recent applications of the framework to presently-relevant research
problems spanning the range from simple 2D to complex 3D systems of
PDEs have been described.

In addition to the presently available capabilities and flexibilities
of the HiFi framework, several development efforts to enhance and
expand the framework's ability to model various idealized,
experimental and naturally occurring physical systems are ongoing or
being planned for the near future.

Implementation of the generalized PBP method to precondition the
linear systems resulting from the implicit advance of PDEs spatially
discretized in the weak form using the spectral element basis set
is one of the ongoing development efforts.  When completed, it is
projected that PBP will allow HiFi to scale to tens of thousands of
processors and beyond.  Furthermore, it will at least halve the amount
of memory presently required to run a given HiFi simulation.

\begin{figure}[htp]
  \center{\includegraphics[width=12cm]{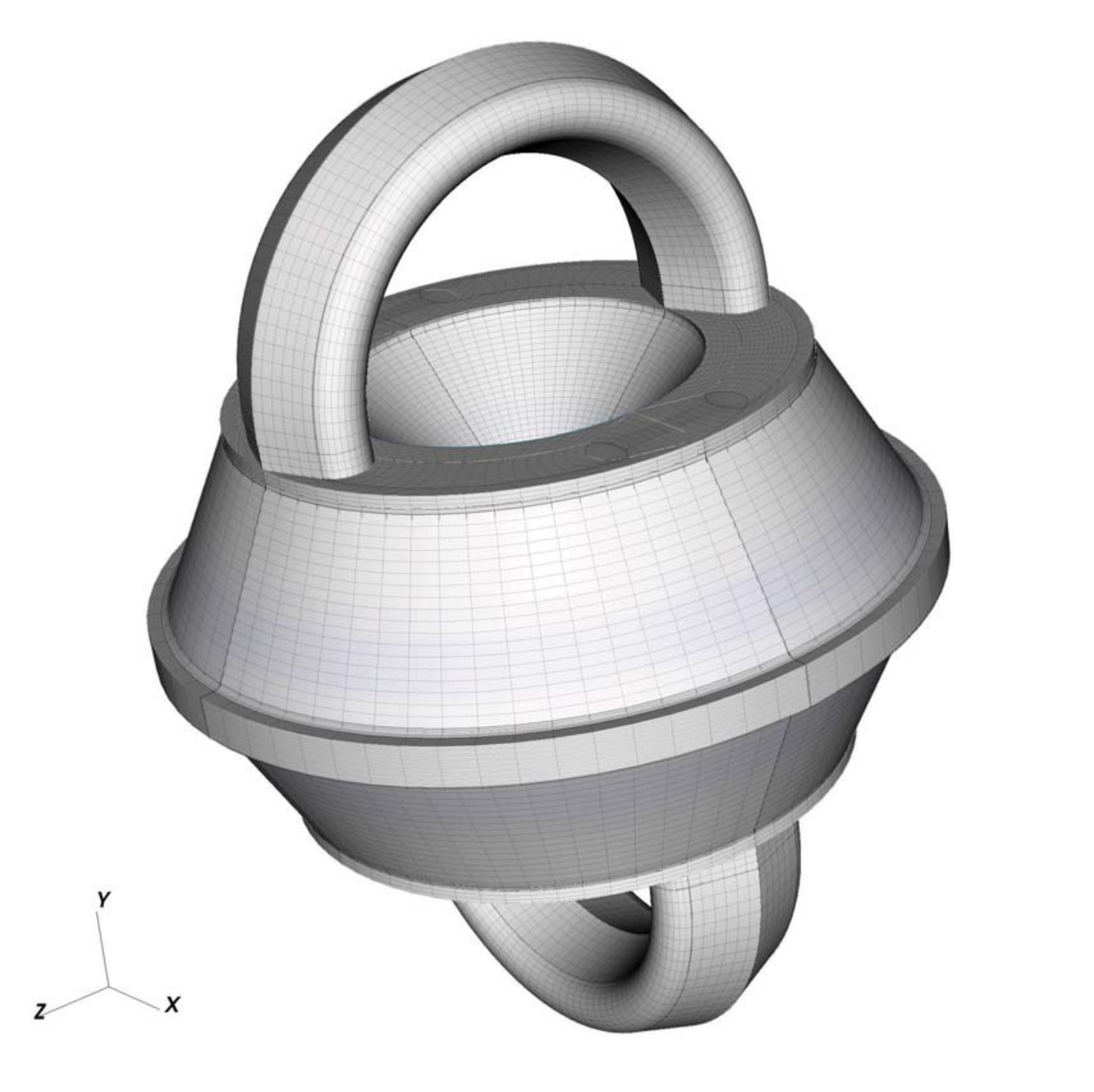}}
  \caption{Multi-block computational grid for a planned extended MHD
    simulation of a HIT-SI experiment\cite{Lowrie11b}.}
  \label{fig:HITSI}
\end{figure}
Another significant and very recent addition to the HiFi toolbox is
the semi-structured grid capability.  Figure~\ref{fig:HITSI} shows an
example of the computational grid composed of several structured grid
blocks that has been successfully used for preliminary testing using
the 3D anisotropic heat conduction equation and the 3D MHD system of
PDEs\cite{Lowrie11b}.  The goal of this development effort has been to
enable HiFi simulations on computational grids of arbitrary
three-dimensional geometry and topology.  In the future, this may
include the ability to use spectral elements of different $n_p$ order
in the different parts of the computational domain, as well as
$n_p$-adaptation.

HiFi is an open-source development project and has been released under
a BSD-style license.  Latest information about the HiFi framework can be found at \href{url}{http://hifi-framework.webnode.com/hifi-framework/}, with verified versions of the framework available to the greater scientific research community upon request.
\section*{Acknowledgements}

This work was supported, in part, by the U.S. Department of Energy and
the Office of Naval Research.  To date, HiFi development has taken
place, in chronological order, at the following institutions: Los
Alamos National Laboratory, Princeton Plasma Physics Laboratory,
University of Washington, and U.S. Naval Research Laboratory.  We
gratefully acknowledge helpful discussions with and occasional coding
contributions from L.~Chac\`{o}n, S.C.~Jardin, M.~Sato,
U.~Shumlak, A.N.~Simakov, C.R.~Sovinec and X.Z.~Tang.

\bibstyle{aip}
\bibliography{Bibl}
\end{document}